# Direct observation of structural heterogeneity and tautomerization of single hypericin molecules

Quan Liu,[1,2] Frank Wackenhut,[1,*] Liangxuan Wang,[1,4] Otto Hauler,[1,3] Juan Carlos Roldao,[4] Pierre-Michel Adam,[2] Marc Brecht[1,3], Johannes Gierschner[4] and Alfred J. Meixner[1,*]

[1] Institute of Physical and Theoretical Chemistry, Eberhard Karls University Tübingen, Auf der Morgenstelle 18, 72076 Tübingen, Germany

[2] Laboratoire Lumière, nanomatériaux & nanotechnologies – L2n and CNRS ERL 7004, Universitéde Technologie de Troyes, 10000 Troyes, France

[3] Reutlingen Research Institute, Process Analysis and Technology (PA&T), Reutlingen University, Alteburgstraße 150, 72762 Reutlingen, Germany

[4] Madrid Institute for Advanced Studies, IMDEA Nanoscience, C/Faraday 9, Ciudad Universitaria de Cantoblanco 28049, Madrid, Spain

**Abstract**：Tautomerization is a fast chemical reaction where structures of the reactants differ only in the position of a proton and a double bond. Tautomerization often occurs in natural substances and is a fundamental process in organic- and biochemistry. However, studying the optical properties of tautomeric species is challenging due to ensemble averaging. Many molecules, such as porphines, porphycenes or phenanthroperylene quinones, exhibit a reorientation of the transition dipole moment (TDM) during tautomerization, which can be directly observed in a single molecule experiment. A prominent phenanthroperylene quinone is hypericin showing antiviral, antidepressive, and photodynamical properties. Here, we study single hypericin molecules by using confocal microscopy combined with higher order laser modes. Observing abrupt flipping of the image pattern allows to draw conclusions about the coexistence of different tautomers and their conversion path. Time-dependent density functional theory calculations show that hypericin is cycling between the four most stable tautomers. This approach allows to unambiguously assign a TDM orientation to a specific tautomer and enables to determine the chemical structure in situ. Additionally, tautomerization can not only be observed by the image pattern orientation, but also as intermittency in the fluorescence emission of a single molecule. Time correlated single photon counting enables to determine the excited state lifetimes of the hypericin tautomers. Our approach is not only limited to hypericin, but can be applied to other molecules showing a TDM reorientation during tautomerization, helping to get a deeper understanding of this important process.

## INTRODUCTION

Tautomerization is the transition between two constitutional isomers of one and the same molecule and could be exploited as a switch in molecular scaled devices[1,2], for sensing[3,4] or influencing the chemical behavior[5,6] and is ubiquitous in nature[7,8]. The most common example is the migration of at least one proton, which occurs e.g. in porphycenes, DNA base pairs or phenanthroperylene quinones. Numerous theoretical studies have been focused on the structural heterogeneity of an ensemble and the mechanisms of tautomerization.[9] However, studying individual tautomeric species is experimentally very challenging, since they usually cannot be isolated due to the similarity of their physical and chemical properties.[10] In a conventional experiment, the properties of the individual tautomeric species are hidden in the ensemble. However, single molecule fluorescence spectroscopy can reveal molecular dynamics and some examples reported in literature are molecular diffusion[11], spectral diffusion[12,13], triplet lifetime variations[14] and intramolecular proton transfer[15]. An approach to directly observe tautomerization is to image the reorientation of the transition dipole moment (TDM) with higher order laser modes, which has already been applied to investigate the double proton transfer in porphycenes.[15] Here, we employ this technique to hypericin, which represents a crucial step forward, both from its biochemical relevance as well as from the challenging photo response. Hypericin is a natural drug in *Hypericum perforatum* (St. John's wort) that is very promising in the treatments of depressive, neoplastic, tumor and viral infections[16]. These medical applications are based on the phototoxic reaction of hypericin. As one of the strongest natural photosensitizers, hypericin is suitable for photodynamic therapy applied for cancer treatment[16–19]. Hence, understanding the interaction of light with hypericin is of scientific interest and is crucial for a successful medical application. This interest is reflected in numerous spectroscopic studies covering the optical properties of hypericin in an ensemble[20–28]. These studies investigate the influence of different solvents and pH on the fluorescence[21] and phosphorescence emission[22,23], fluorescence lifetime[24], and SERS spectrum[25], excitation wavelength dependence of the Raman and SERS spectrum[26,27] and excited-state proton transfer[28].

However, a major challenge is the complexity of the dynamics of hypericin, since it can undergo excited and ground state tautomerization, de-/protonation, conformational (torsional) transitions, as well as possible association (dimerization)[9,29]. Hence, it is difficult to pin down the origin of a specific change in the photophysical properties, which is important to explain the variation of viricidal activity[30]. For the tautomerization process, semi-empirical simulations[9,29] show ten possible tautomers of hypericin. The most stable species is the tautomer $Q_{7,14}$ (carbonyl group at position 7 and 14) and the migration of one proton results in three metastable mono-tautomerized species. The remaining six double-tautomerized species are less likely to be present due to larger activation barriers and the requirement of two-step interconversion. Additionally, these interconversions can be further classified into intra- and inter- molecular tautomerization depending if the proton is migrating within the same hypericin molecule or needs to be offered by the surrounding solvent cage. Importantly, the TDM of the tautomers differ in strength and orientation, as shown by our time-dependent density functional theory (TD-DFT) calculations. In fact, (TD)DFT permits a reasonable description of isomer energies[31], transition states[32] and TDM orientation[33], and thus enables a detailed correlation with experimental studies. Here, we monitor the tautomerization by optical imaging with higher order laser modes.[34–36] This technique allows to directly image the orientation of the TDM and tautomerization can be observed by a flipping of the image pattern orientation. The combination of this technique with TD-DFT allows to unambiguously assign a certain orientation of the TDM to a specific tautomer. Therefore, the tautomerization state, and hence the chemical structure, of a single hypericin molecule can be determined in-situ and conclusions on the coexistence of different tautomers of hypericin can be drawn.

## MATERIAL AND METHODS

**Sample preparation**: Hypericin (Burg-Apotheke, Königstein) was dissolved in ethanol (Uvasol, Merck) and diluted to a concentration of $10^{-9}$ M for single molecule fluorescence experiments. Poly (vinyl alcohol) (PVA, Sigma-Aldrich) was dissolved in triply distilled water, and 2 µL of hypericin solution was added to a 2 wt. % PVA solution to obtain the required concentration. All solutions were stocked in the dark and at 4 °C to minimize molecule bleaching. For single molecule measurement, coverslips were cleaned in chromosulfuric acid solution for 4 hours, rinsed with triply distilled water and finally dried in a nitrogen flow. For single molecule imaging, time trace, and antibunching measurements, 20 µL of the $10^{-9}$ M hypericin/PVA solution was spin coated (6k rpm, 30 s) on a coverslip and the resulting polymer film has a thickness of about 80 nm according to AFM measurement.

**Single molecule spectroscopy**: Fluorescence images of single hypericin molecules were recorded with a home-built scanning confocal microscope[36] with a 530 nm pulsed laser (pulse duration < 100 ps, 20 MHz) as excitation source. The azimuthally polarized (APDM) and radially polarized (RPDM) doughnut modes are generated by a commercial mode converter (Polarization Converter, ARCoptix). The excitation laser (1.2 µW in front of the objective) was focused on the sample by a high numerical aperture (NA = 1.46, Carl Zeiss) oil objective lens. The fluorescence signal was collected by the same objective lens and sent to two avalanche photodiodes (APDs, SPCM-AQR-14, PerkinElmer) via a 50/50 beam splitter. APDs are connected to a time-correlated single photon counting module (TCSP, HydraHarp 400, PicoQuant).

**TD-DFT**: DFT calculations were performed to optimize the conformational isomers of the most stable tautomer $Q_{7,14}$ by torsions of the aromatic backbone (i.e. $^P$M-Q, $^B$M-Q; being isoenergetic to $^P$P-Q and $^B$P-Q, respectively) and the conformers generated by H-flip at the 3-, 4-positions; minima were confirmed by the absence of imaginary frequencies. Transition states between the minima were calculated, giving one imaginary frequency, which reflect the path of the conformational change between the minima. Then, all possible ten tautomers were optimized; it is noted that the different tautomers are not characterized by the position of the carboxyl-groups, but also by the orientation of the adjacent hydroxyl groups, so that the tautomerization occurs in a two-step process in some cases. Finally, the orientation of the transition dipole moment for each tautomer between ground and first excited state ($S_0 \rightarrow S_1$) were calculated by single point TD-DFT calculations. All calculations were done employing the B3LYP functional and the 6-311G* basis set, as described in the Gaussian16 program package[37].

## RESULTS AND DISCUSSION

**Time-Dependent Density Functional Theory of Hypericin Tautomers**. Hypericin displays a rich structural variety in terms of conformational (i.e. torsional, see Figure S1 and S2) and constitutional isomerization.[9] According to Ref. [9], four stable conformers with different dihedral deformation, so called butterfly (B) and propeller (P) conformers are observed, and our DFT-calculations of the dihedral angles of the P-forms in the free molecule (28º and 32º; see SI) agree reasonably with X-ray crystallographic study (19º and 32º).[38] For the constitutional isomers, ten tautomers were identified, whose relative stability was estimated earlier by force field calculations.[28] As more accurate energies are required for the envisaged correlation with experiment, the tautomers were here optimized by DFT (Figure 1 and Table S1), which differ significantly from the earlier result in Gibbs free energies and relative stability. Following the established nomenclature[29], labelled numbers (1-14) are used to describe the position of carbonyl groups; for instance $Q_{7,14}$ denotes the most stable tautomer with the carbonyl groups in positions 7 and 14.

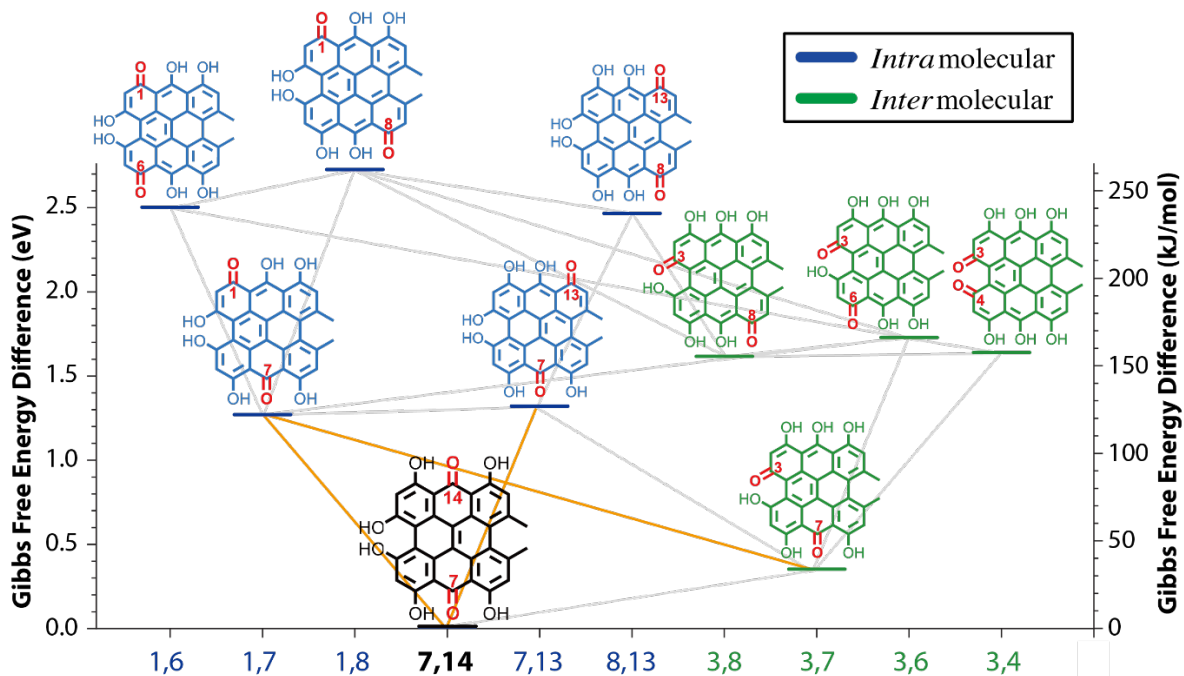

Figure 1. Gibbs free energy (calculated by DFT) diagram of the tautomers of hypericin overlaid with the tautomerization transition network. Two classes of tautomers, accessible via intra- and inter-molecular proton transfer, are sketched in blue and green, respectively. The connected grey lines between two tautomers represent the possible proton transfer. The orange lines mark the transitions observed in Figure 4.

It is noted that the tautomers fall in two classes; starting from the most stable tautomer $Q_{7,14}$ (black in Figure 1), five are accessible by *intra*molecular proton transfer (blue in Figure 1) via one-step ($Q_{1,7}$, $Q_{7,13}$) or two-step tautomerization ($Q_{1,6}$, $Q_{1,8}$, $Q_{8,13}$). Realization of the other tautomers from $Q_{7,14}$ require *inter*molecular proton transfer (drawn in green) from the solvent via one-step ($Q_{3,7}$) or two-step ($Q_{3,4}$, $Q_{3,6}$, $Q_{3,8}$) tautomerization. Thus, starting from the most stable $Q_{7,14}$, only three structures ($Q_{1,7}$, $Q_{7,13}$, $Q_{3,7}$) are accessible via single proton transfer and are relatively low in energy with $\Delta E \leq 1.3$ eV and the activation barriers are larger than thermal energy $kT$ [1]. Therefore, quantum tunneling[39] must play a dominant role for the tautomerization of hypericin. The other tautomeric species exhibit even higher energy barriers and require double proton transfer; hence, they are unlikely to occur under normal experimental conditions, and can be safely neglected in the following considerations. The significant change of the carboxyl group position during tautomerization is accompanied with an electronic redistribution; hence, the $S_0$-$S_1$ electronic transition dipole moment (TDM) has a different orientation for the four tautomers. It is noted that the different conformations mentioned above do not cause any change of the TDM orientation according to the TD-DFT calculations, since the out-plane angles of the TDMs are small and a conformational transition does not change the in-plane symmetry. For this reason, conformational changes cannot be distinguished by imaging with higher order laser modes. Furthermore, the computed energy barrier for the conformational transitions are several times larger compared to tautomerization[9,29] so that conformational transitions can be excluded in our experiment.

---

[1] The activation barriers for tautomerization were calculated with DFT, which involve a complex two-step process of proton abstraction and flipping; details of this study will be given elsewhere

**Single Molecule Microscopy and Emission Dynamics of Hypericin.** TDMs calculated by TD-DFT are shown by the colored arrows (blue, yellow, green, cyan) in Figure 2a. The exact three-dimensional orientation and relative strength of the TDMs is shown in Figure 2b together with the molecular structure. The molecular orientation is chosen in a way that the molecular coordinate system xyz is overlapping with the scanned experimental image coordinate system XYZ, where the X-Y plane is perpendicular to the optical axis of the microscope with Z = 0 being the center of the focus. The color of the TDM vectors corresponds to the tautomers in Figure 2a. All the TDMs lie almost in X-Y plane and the out-of-plane angles are less than 2°. For this orientation, the largest angle difference between TDMs from $Q_{1,7}$ and $Q_{7,13}$ is 24° in the X-Y plane.

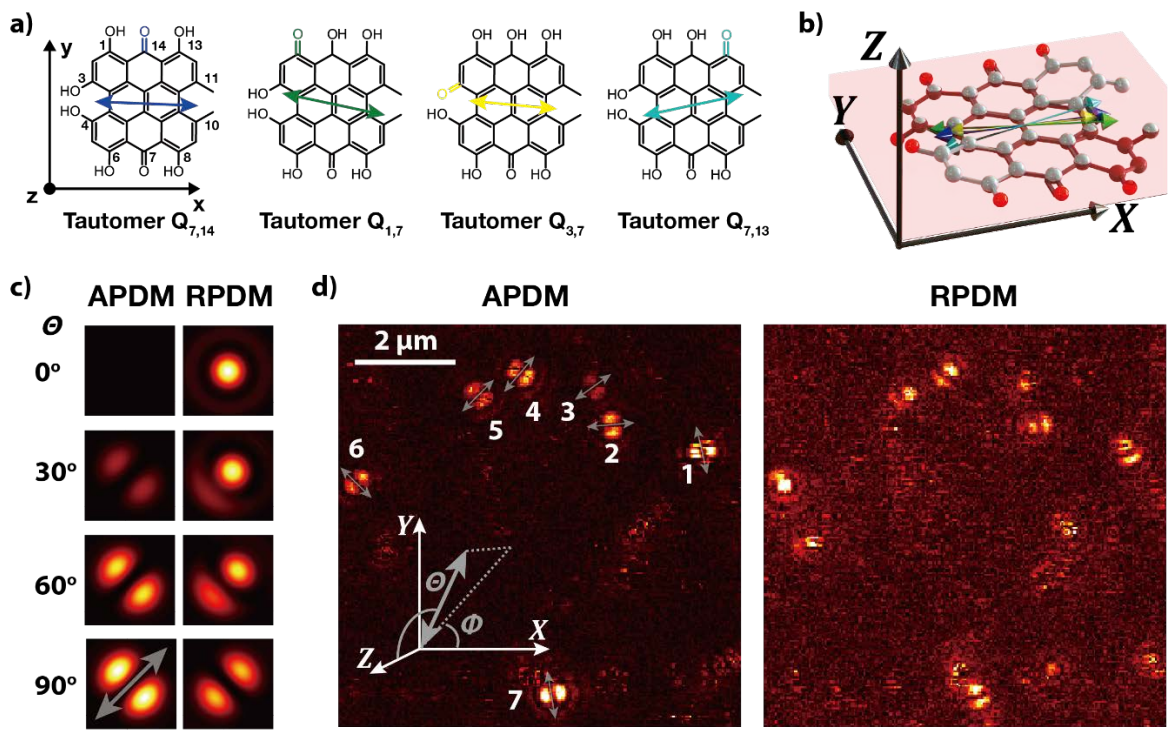

Figure 2. (a) Chemical structure of hypericin tautomer $Q_{7,14}$ and three other most stable tautomeric species. Colored arrows indicate the orientation of $S_0$-$S_1$ electronic transition dipole moments (TDM). The molecular coordinate system xyz is fixed on the molecule. (b) Three-dimensional molecular structure and TDMs when molecular coordinate system xyz is overlapped with image coordinate system XYZ. (c) Simulated fluorescence excitation patterns of different TDM orientations. The polar angle $\Theta$ and azimuthal angle $\Phi$ are defined in the image coordinate system XYZ shown in d. (d) Exemplary confocal fluorescence image of separated single hypericin molecule embedded in a PVA matrix acquired with azimuthal (APDM, left) and radial (RPDM, right) polarization, respectively. Scanning range is 8μm × 8μm (240 pixels × 240 pixels, 5 ms per pixel). Grey arrows illustrate the orientation of the TDM in the X-Y plane.

In order to investigate the tautomerization, we use an excitation laser beam with azimuthal (APDM) and radial (RPDM) polarization[34]. The three-dimensional orientation of the TDM is defined by the azimuthal $\Phi$ and polar $\Theta$ angle in the image coordinate system XYZ (0° ≤ $\Phi$ ≤ 360°, 0°≤ $\Theta$ ≤ 180°). Figure 2c shows simulated excitation patterns of a single TDM with a fixed in-plane angle $\Phi$ = 45° and an out-of-plane angle $\Theta$ ranging from 0° to 90°. Variations of the in-plane angle $\Phi$ can be observed by a rotation of the respective pattern. Figure 2d shows an image of single hypericin molecule acquired with the APDM and all patterns have a double lobed shape. The APDM is exclusively transversally polarized in the X-Y plane in the focus, hence the pattern orientation directly displays the orientation of the TDM in the X-Y plane (indicated by the grey arrows, referred to in-plane dipole)[34–36]. Excitation with the RPDM results in patterns ranging from dot-like to asymmetric double lobe to symmetric shapes. These different pattern shapes are caused by the interaction of transversal (in-plane) and longitudinal (out-of-plane) polarizations in the focus of the RPDM with the three dimensionally oriented TDM (see Figure 2c). This allows to determine the full three-dimensional orientation of the TDM. The dark lines in the image patterns of molecule **1** and **4** in Figure 2d show blinking. These molecules were in a dark state for a certain time, which is a first indication that a single molecule is observed. The fluorescence patterns in Figure 2d prove that the single hypericin molecules have a random, but fixed, orientation in the PVA matrix. As illustrated in Figure 2a and 1b, the four tautomers of hypericin have a different orientation of the TDM and a transition between these tautomers can be observed by a change of the image pattern orientation[15,40]. Figure 3a shows three consecutive scan images of the same two single hypericin molecules.

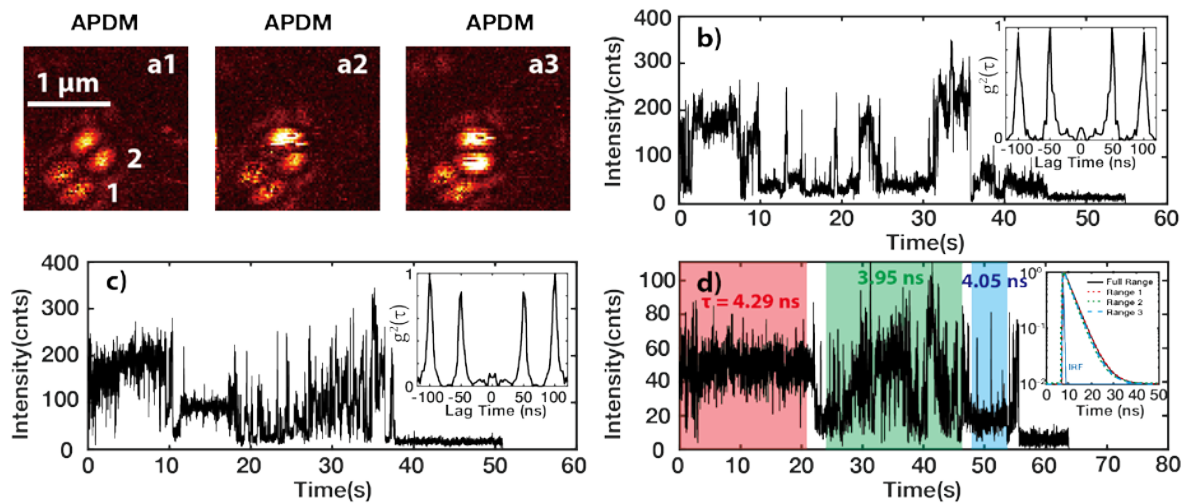

Figure 3. (a) Three consecutive images acquired with the APDM at the same spatial position. Scanning range is 2 µm × 2 µm (80 pixels × 80 pixels, 5 ms per pixel). (b) and (c) Two intensity time traces of two different single hypericin molecules with a binning time of 10 ms in coincidence mode. Insets show the second-order intensity correlation function of the respective time traces. (d) Intensity time trace of a single hypericin molecule with a binning time of 10 ms in lifetime mode. Inset figure is normalized exponential decay of different emission states (red, green, blue). Single exponential fitting reveals that lifetime range from 3.95 ns to 4.29 ns.

The double lobe pattern orientation and therefore the TDM orientation of molecule 1 has a fixed orientation in the image series in Figure 3a. However, the orientation of the TDM of molecule 2 does rotate by 40° from a1 to a3. Especially in a2, the transient dynamic of the tautomerization transition manifests itself as a sudden flip of the excitation pattern orientation and intensity. These images suggest, that the tautomerization transition process can also be observed from the fluorescence intensity emitted by the molecule. Figure 3b, c show intensity time traces of two different hypericin molecules. The total number of collected photons from each hypericin molecule is about $3.4 \times 10^5$. Both time traces exhibit multiple intensity steps, which is usually an indication that more than one molecule is present in the excitation focus. However, both molecules show clear antibunching in the second-order intensity correlation function (insets in Figure 3b and c), which proves that the emission does indeed originate from a single molecule. Same multi-step behavior is also observed in the intensity time trace acquired during lifetime measurements, as shown in Figure 3d. Remarkably, there is no significant lifetime difference between the colored areas (4.29 ns for red, 3.95 ns for green, 4.05 ns for blue), even though the fluorescence intensity is different. This indicates that a grey state[41], i.e. an emission state with a different radiative rate, or association/dissociation[24] of hypericin are not the causes of the intensity fluctuations. We assume, that the quantum yield of the same molecule and the collection efficiency of the system are the same for two reasons, first, there is no difference in fluorescence lifetime and second we use the same confocal scanning microscope for all the measurements. Therefore, the intensity fluctuation can be assigned to a tautomeric transition, since the orientation of electronic TDM relative to the excitation field changes and the excitation rate $\gamma_{exc}$ will be different, as it is proportional to the absolute square of the scalar product of the dipole moment and the electric field ($\gamma_{exc} \propto |\vec{P} \cdot \vec{E}|^2$). Hence, reorientation of the TDM by tautomerization will cause abrupt fluorescence intensity changes and lead to multiple intensity steps observed in Figure 3b-d. From this temporal dynamic of the intensity time traces, we can conclude that a single hypericin molecule can stay in the same tautomeric state for up to tens of seconds. This is consistent with the observation of a flipping double lobe pattern in Figure 3a, which can only be observed when the tautomerization transition is slower than the image acquisition time.

**Tautomerization of Hypericin observed by Imaging with Higher Order Laser Modes**. In order to further verify that the observed flipping of the TDM is caused by a tautomerization transition, we acquired scan images of the same spatial region with APDM and RPDM until photobleaching of the molecules occurred. An exemplary image series of the same spatial area is shown in Figure 4.

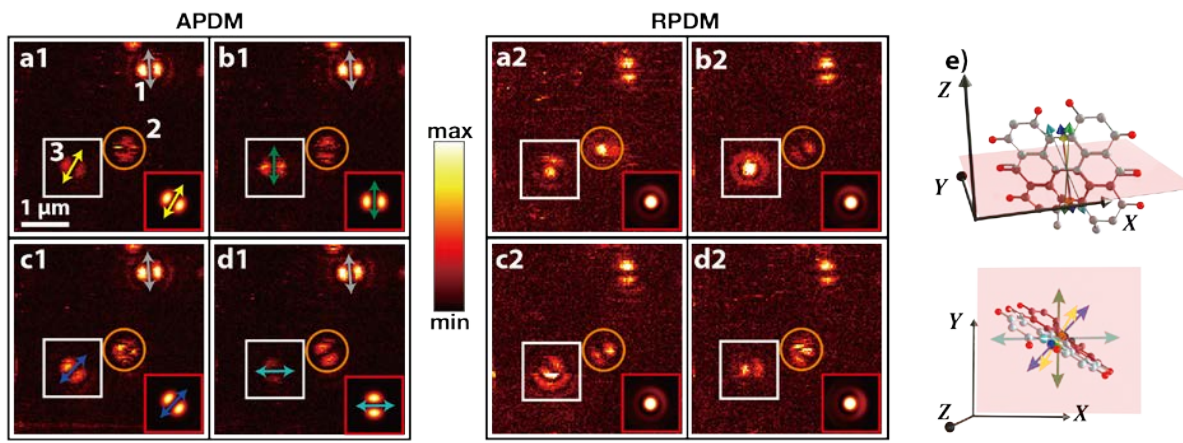

Figure 4. Eight exemplary images recorded at the same spatial position. Scanning range is 4 µm × 4 µm (100 pixels × 100 pixels, 5 ms per pixel). (a1-d1) are excited with APDM and (a2-d2) are excited with RPDM. Inset images in red squares (1.2 µm × 1.2 µm) are simulation results. Colored arrows correspond to the TDMs of the tautomers in Figure 2. (e) Orientation of molecule 3 in the image coordinate system XYZ together with the electronic transition dipoles. The top shows the orientation of the transition dipoles in the image coordinate system XYZ, while the in-plane projection is illustrated on the bottom. Colored arrows correspond to the scaled projection length and the orientation of the transition dipoles in the X-Y plane.

Fluorescence patterns of three single molecules can be observed in Figure 4 and the images are arranged in pairs (a1-a2, b1-b2, …) according to the transition dipole orientation of molecule 3. The pattern orientation of molecule 1 is stable during all measurements in Figure 4, hence this molecule is always in the same tautomeric state and serves as reference for molecule 2 and 3. Molecule 2 has a ring like pattern shape in a1 and c1, showing that the change of the pattern orientation is faster than the image acquisition time. For molecule 3 (marked white square) four pairs of different excitation patterns can clearly be distinguished representing four different TDM orientations that are consistent with the four most stable tautomers of hypericin (see Figure 2a). The whole image series consists of 18 consecutive scans (see supporting information, Figure S3), but molecule 3 does not show flipping in every image and we represent only the ones with a change of the pattern orientation. In order to obtain the three-dimensional spatial orientation of molecule 3, the excitation patterns in image a1 was first fitted with 2D Gaussian functions with a self-written Matlab script. This gives an orientation of the double lobe pattern with an azimuthal angle $\Phi \approx 60°$ in a1/a2. Numerical simulations[42,43] were performed to compute a series of image patterns with fixed azimuthal angle $\Phi$, and varied polar angle $\Theta$ under RPDM excitation. Comparison between simulated image patterns and the experimental image in Figure 4) a2) allows to determine the polar angle to be $\Theta \approx 165°$. Once the three dimensional orientation of one TDM is known in the image coordinate system XYZ, the rotation matrix **R** from the image coordinate system to the molecular coordinate system xyz can be solved (for details see supporting information)[44]. However, at this stage it is unknown which tautomer is observed in Figure 4) a2) and without a priori information about the relationship between the tautomers, there are four possible solutions of **R**.

The spatial orientation of the other three tautomers in the image coordinate system can be obtained by applying the rotation matrix **R** to the TDM orientations in the molecular coordinate system obtained from the TD-DFT calculations. After this rotation, the actual polar $\Theta$ and azimuthal $\Phi$ angles for all four dipoles in the image coordinate system are known and the only possible orientation of molecule 3 agreeing with the experimental results is shown in Figure 4e. The corresponding angles are summarized in Table S1 in the supporting information. These TDM orientations can be used to calculate the respective image pattern in the image coordinate system, which is shown in the inset images in red squares. Please be aware that only the first image pattern in a1 is fitted to determine the orientation of one TDM and the pattern orientations, shown in the red squares of Figure 4, of the other three tautomers is purely obtained from the relative TDM orientations obtained from TD-DFT calculations. It is worth to mention that the pattern of molecule 3 in Figure 4) c2) is an overlap of the tautomer $Q_{7,14}$ and $Q_{1,7}$ and tautomerization occurred during the scan. There is an excellent agreement between the experimental data and the TDM orientations obtained by TD-DFT calculations, which confirms that the observed flipping of the image patterns is caused by tautomerization and that we can identify which tautomer is observed in each image. The corresponding transition pathways observed in Figure 4 are marked as orange lines in Figure 1. Another example is shown in Figure S4 of the supporting information and the corresponding angles are given in Table S2. Please be aware, that the polar angle of the molecule 3 shown in Figure 4 is larger than 90° (the methyl groups are pointing in negative z-direction), while it is smaller than 90° for the molecule in Figure S4 (the methyl groups are pointing in positive z-direction). This shows that it is possible to determine the full 3D orientation of a single molecule.

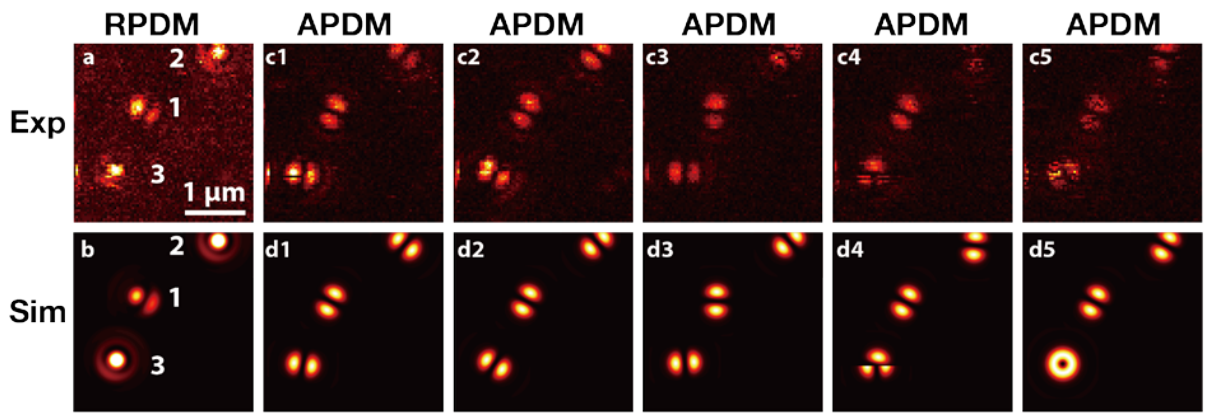

Figure 5. Consecutive scans of three single hypericin molecules. Scanning range is 3 µm × 3 µm (80 pixels × 80 pixels, 5 ms per pixel). Experimental (a, c) and simulated (b, d) fluorescence excitation patterns acquired at same spatial position with (a, b) RPDM and (c, d) APDM excitation. The images are acquired continuously within 10 minutes.

Figure 5 shows experimental excitation patterns of three different single molecules acquired with the RPDM (a) and APDM (c1-5). Molecule 1 is nearly oriented parallel to the X-Y plane with a polar angle of ~60°, as can be immediately be seen from RPDM excitation patterns. Molecule 2 and molecule 3 are nearly perpendicular to the X-Y plane with the polar angles of 14° and 11°, respectively. Molecule 1 shows only small variations of the pattern orientation, hence the polar and azimuthal angle are directly obtained by fitting the excitation patterns. The rotation from the molecular to the image coordinate system described above is applied to molecule 2 and 3. The corresponding simulation results for all three molecules are presented in the panels label as b for the RPDM and d1-d5 for the APDM. The azimuthal and polar angles are summarized in Table S2. Interestingly, molecule 1 only shows small in-plane angle variations of less than 30°, while large in-plane angle variations of 124° and 109° are observed for molecule 2 and molecule 3. Since the angle difference of the TDMs between different hypericin tautomers is smaller than 24°, it is only possible to have large in-plane angle variation when the dipole moments are about to vertical. This means, that large in-plane angle variations of the image pattern caused by tautomerization can only be observed for molecules with a TDM orientation nearly perpendicular to the X-Y plane. Molecule 3 shows a sudden transition from $Q_{7,14}$ to $Q_{7,13}$ in c4, and shows a fast tautomerization rate in c5 resulting in a ring like pattern shape. Another example of molecules nearly oriented parallel to the X-Y plane is presented in the supporting information Figure S5. Consistently, only small variations of the pattern orientation can be observed for these molecules. This does not mean that there is no tautomerization, but the resulting pattern reorientation is smaller and is more difficult to be observed.

**Tautomerization Mechanism of Hypericin**. The results presented here show that the same hypericin molecule cycles between the four most stable tautomers and that imaging with higher order laser modes is suitable to directly image this three-dimensional structural change. The combination of confocal microscopy with higher order laser modes and TD-DFT calculations gives the first experimental prove for the coexistence of different tautomers of hypericin, and the tautomeric species can unambiguously be identified in each experimental image. In the present observation time window, the transition between two tautomeric species appears as a fast sudden event; afterwards, the single hypericin molecule can stay in the same tautomeric state up to tens of seconds. These results show that fluorescence intensity fluctuations of a single hypericin molecule on a time scale of several seconds can been assigned to tautomerization. Additionally, these results shed light on the mechanism of tautomerization of hypericin. The energy of an excitation photon (2.34 eV) is larger than the barriers between the tautomers as shown in Figure 1. This suggests that hypericin tautomerization could in principle be induced by excitation photons, as was discussed intensively in the past.[20,28] However, under the current experimental conditions, if the transitions are initiated by the excitation laser, the transition should always occur as a sudden flipping during image acquisition like in Figure 3) a2). In most cases, the molecule stays in the same tautomeric state during an entire scan and the same molecule is in a different state at the beginning of next scan image. For example, only 2 out of 18 consecutive scans of molecule 3 in Figure 4/S1 show a sudden reorientation of the pattern during image acquisition. This indicates that most transitions occur without optical excitation and hence in the electronic ground state. As mentioned above, the high barriers for tautomerization imply that quantum tunneling must play a dominant role for the tautomerization of hypericin. It is finally noted that hypericin tautomerization modifies the position of labile protons, which determine its ability to acidify its environment and are related to the viricidal activity of hypericin[28,45]. Understanding the equilibrium of hypericin tautomer heterogeneity can be important to explain the variation of viricidal activity.

## CONCLUSION

In conclusion, we presented the in-situ observation of single molecule tautomerization dynamic, proving the coexistence of the four most stable tautomers of hypericin. Additionally, TD-DFT allows to determine relative orientation of the TDMs of the tautomeric species. This enables to determine the three-dimensional spatial orientation of the molecule and to identify the tautomeric state observed in a specific image. The calculated relative orientation of the TDMs also explains why large angle variations of image pattern are only observed, when the molecule is oriented perpendicular to the image plane. Additionally, this work demonstrates the usefulness of confocal microscopy combined with higher order laser modes to investigate the temporal dynamics of chemical reactions. In the case of hypericin, we find large fluctuation of the tautomerization rates of different molecules, depending on their local environment. This effect can be exploited to optimize the environment in drug delivery systems for the best performance in photodynamic therapy. The method of confocal microscopy with higher order laser mode has the advantage that it can be synchronized with additional techniques, like spectroscopic or fluorescence lifetime imaging (FLIM), to access more information of tautomers and temporal dynamics ranging from picoseconds to seconds. The same technique can easily be applied to observe temporal dynamics of other molecules or nanoparticles[46].

## AUTHOR INFORMATION

**Corresponding Author**

**Frank Wackenhut** − *Institute of Physical and Theoretical Chemistry, Eberhard Karls University Tübingen, 72076 Tübingen, Germany;* orcid.org/0000-0001-6554-6600; Email: frank.wackenhut@uni-tuebingen.de

**Alfred J. Meixner** − *Institute of Physical and Theoretical Chemistry, Eberhard Karls University Tübingen, 72076 Tübingen, Germany;* orcid.org/0000-0002-0187-2906; Email: alfred.meixner@uni-tuebingen.de

**Authors**

**Quan Liu** − *Institute of Physical and Theoretical Chemistry, Eberhard Karls University Tübingen, 72076 Tübingen, Germany; Laboratoire Lumière, nanomatériaux & nanotechnologies − L2n and CNRS ERL 7004, Université de Technologie de Troyes, 10000 Troyes, France*

**Liangxuan Wang** − *Institute of Physical and Theoretical Chemistry, Eberhard Karls University Tübingen, 72076 Tübingen, Germany; Madrid Institute for Advanced Studies, IMDEA Nanoscience, C/Faraday 9, Ciudad Universitaria de Cantoblanco 28049, Madrid, Spain*

**Juan Carlos Roldao** − *Madrid Institute for Advanced Studies, IMDEA Nanoscience, C/Faraday 9, Ciudad Universitaria de Cantoblanco 28049, Madrid, Spain*

**Otto Hauler** − *Institute of Physical and Theoretical Chemistry, Eberhard Karls University Tübingen, 72076 Tübingen, Germany; Reutlingen Research Institute, Process Analysis and Technology (PA&T), Reutlingen University, Alteburgstraße 150, 72762 Reutlingen, Germany*

**Pierre-Michel Adam** − *Laboratoire Lumière, nanomatériaux & nanotechnologies − L2n and CNRS ERL 7004, Université de Technologie de Troyes, 10000 Troyes, France*

**Marc Brecht** − *Institute of Physical and Theoretical Chemistry, Eberhard Karls University Tübingen, 72076 Tübingen, Germany; Reutlingen Research Institute, Process Analysis and Technology (PA&T), Reutlingen University, Alteburgstraße 150, 72762 Reutlingen, Germany*

**Johannes Gierschner** − *Madrid Institute for Advanced Studies, IMDEA Nanoscience, C/Faraday 9, Ciudad Universitaria de Cantoblanco 28049, Madrid, Spain*

**Notes**

The authors declare no competing financial interest.

## ACKNOWLEDGMENT

The authors gratefully acknowledge funding by the German Research Foundation (DFG, ME 1600/13-3) and the Funding of China Scholarship Council (CSC). The work in Madrid was supported by the Spanish Science Ministry (MINECO-FEDER projects CTQ2017-87054, SEV-2016-0686) and by the Campus of International Excellence (CEI) UAM+CSIC. L.W. acknowledges an Erasmus+ grant of the European Commission. L.W. and J.G. thank Begoña Milián-Medina, Valencia.

# Supporting information for:

**Direct observation of structural heterogeneity and tautomerization of single hypericin molecules**

1. Conformation transition

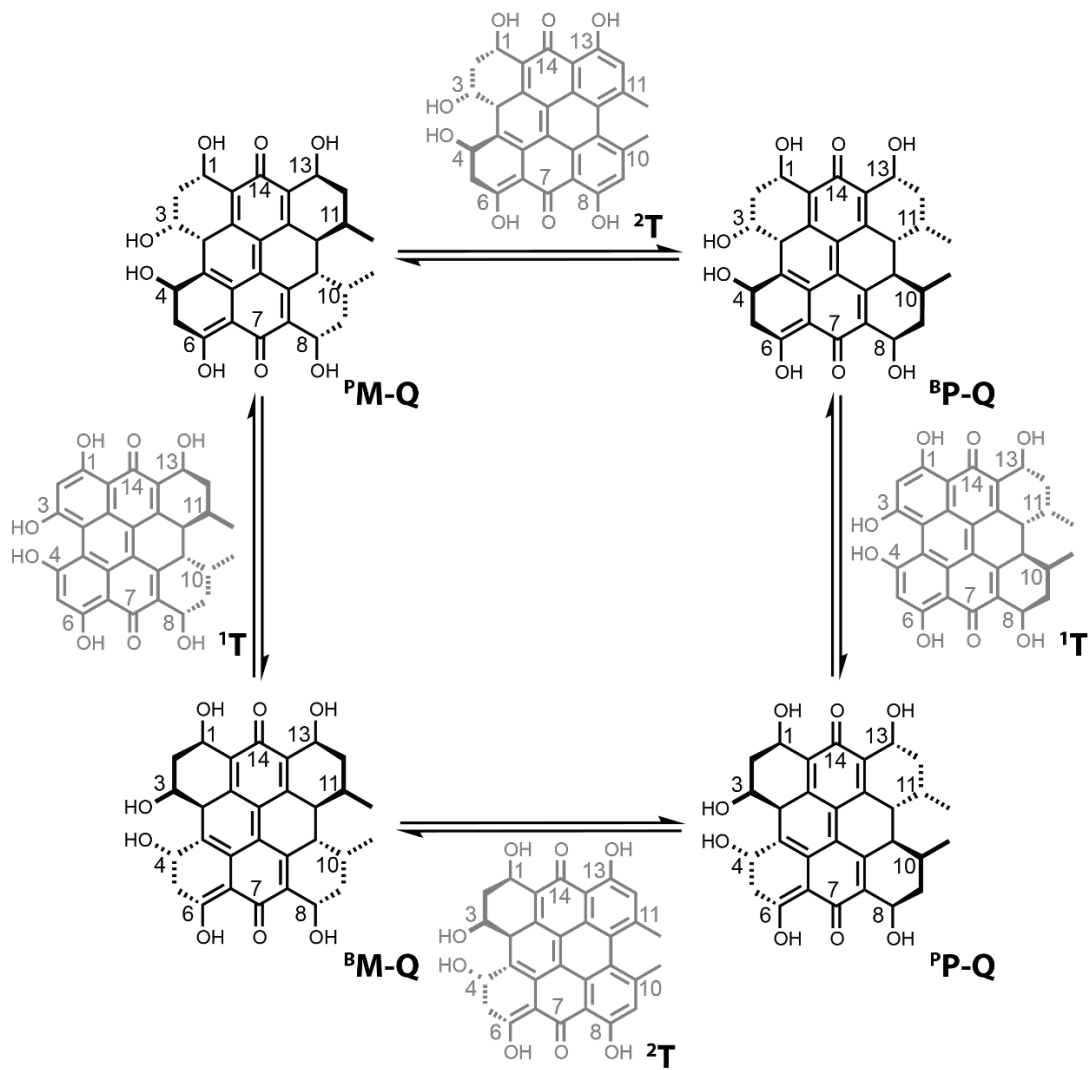

Figure S1 Conformation transition diagram between the propeller ($^P$M-Q and $^P$P-Q) and butterfly ($^B$M-Q and $^B$P-Q) tautomers with the corresponding transition states ($^1$T and $^2$T).

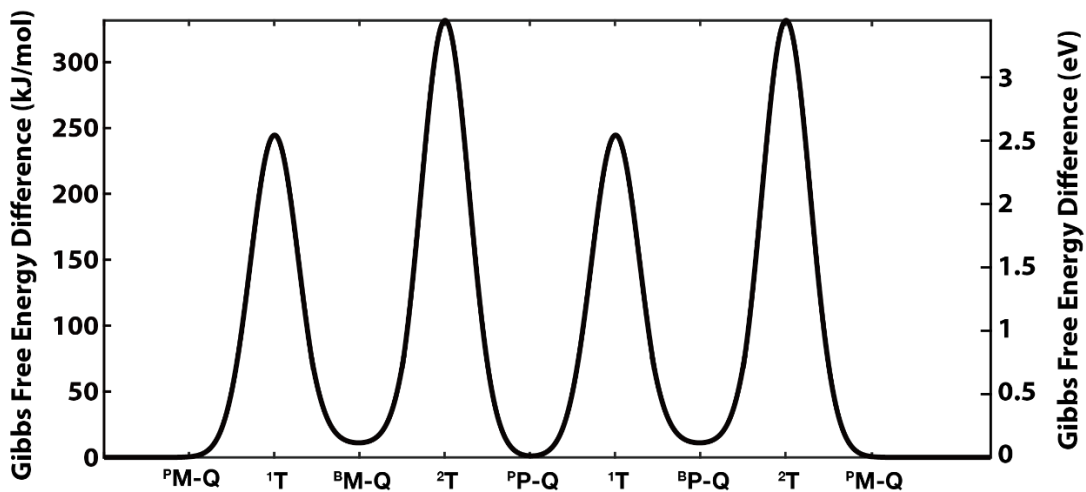

Figure S2: Gibbs free energies (calculated by DFT) of the propeller ($^P$M-Q and $^P$P-Q) and butterfly ($^B$M-Q and $^B$P-Q) tautomers with the corresponding transition states ($^1$T and $^2$T).

## 2. Intra- and inter-molecular transfer

The diagram in Figure 1) shows the Gibbs free energy difference between tautomer $Q_{7,14}$ and the other tautomers, the energy difference is listed in Table S1. Considering the proton transfer distance and if there is a formation of new bond, the tautomers fall in two classes; starting from the tautomer $Q_{7,14}$, five are accessible by *intra*molecular proton transfer in a one-step ($Q_{1,7}$, $Q_{7,13}$) or a two-step tautomerization ($Q_{1,6}$, $Q_{1,8}$, $Q_{8,13}$) or by *inter*molecular proton transfer in a one-step ($Q_{3,7}$) or two-step ($Q_{3,4}$, $Q_{3,6}$, $Q_{3,8}$) tautomerization. Figure 1 shows the corresponding tautomerization network.

We specially note *inter*molecular proton transfer tautomer $Q_{3,7}$, which is not possible to access in an *intra*molecular fashion. The first deprotonation of hypericin occurs at the *bay* hydroxyl groups with a low p$K_a$ of 1.8, and it has been observed in our previous experiment with similar apparatus[1]. Hence, *intra*molecular proton transfer could start via deprotonation at the *bay* hydroxyl groups, and followed by the protonation of *peri* carbonyl group with PVA matrix severs as proton donor.

Table S1 Gibbs free energy difference between tautomer Q$_{7,14}$ and other tautomers

| Tautomer $^P$M-Q | Gibbs Diff. (eV) | Gibbs Diff. (kJ/mol) |
|---|---|---|
| 7,14 | 0 | 0 |
| 7,13 | 1.31 | 125.97 |
| 8,13 | 2.46 | 235.8 |
| 1,6 | 2.49 | 239.34 |
| 1,7 | 1.26 | 121.4 |
| 1,8 | 2.72 | 260.8 |
| 3,4 | 1.63 | 156.46 |
| 3,6 | 1.72 | 165.02 |
| 3,7 | 0.34 | 32.91 |
| 3,8 | 1.61 | 154.22 |

3. **Calculation of rotation matrix R.**

We can define two coordinate systems, the molecular coordinate system **xyz** fixed on center of mass of the molecule, and the image coordinate system **XYZ.** TD-DFT yield the orientation of the transition dipole moment in the molecular coordinate system **xyz**, while images are acquired in the image coordinate system **XYZ**. The rotation matrix[2] **R** between the two coordinate systems is defined as:

$$\begin{bmatrix} X \\ Y \\ Z \end{bmatrix} = R \begin{bmatrix} x \\ y \\ z \end{bmatrix}$$

From DFT calculations, we have obtained the three-dimensional orientation of the transition dipole moments of the four tautomers **d$_1$, d$_2$, d$_3$, d$_4$** defined in the molecular coordinate system **xyz**. From the experimental patterns in different confocal images, it is able to determine the three-dimensional orientation of the transition dipole moments **D$_1$, D$_2$, D$_3$, D$_4$** in the image coordinate system **XYZ** for a particular single molecule. At first, the tautomerization state of the single molecule is unknown in the individual images, but the relative orientations of transition dipole moments **D$_1$, D$_2$, D$_3$, D$_4$** can be obtained from the image series. Assuming the three-dimensional orientation of **D$_1$** is already obtained from image fitting. However, without a priori information about the relationship between **d$_1$, d$_2$, d$_3$, d$_4$** and **D$_1$, D$_2$, D$_3$, D$_4$** we obtain four different rotation matrix **R$_1$, R$_2$, R$_3$, R$_4$**, by solving the following equation:

$$\boldsymbol{D_1} = R\boldsymbol{d}_i, i = 1, \dots, 4$$

This yields four different sets of dipole orientations in **XYZ**, which are **R$_1$(d$_1$, d$_2$, d$_3$, d$_4$), R$_2$(d$_1$, d$_2$, d$_3$, d$_4$), R$_3$(d$_1$, d$_2$, d$_3$, d$_4$), R$_4$(d$_1$, d$_2$, d$_3$, d$_4$)**. Comparing with **D$_1$, D$_2$, D$_3$, D$_4$**, we can determine the correct rotation matrix **R** and identify the tautomerization state in the individual experimental scan images.

4. **Full series of successive images shown Figure 4**

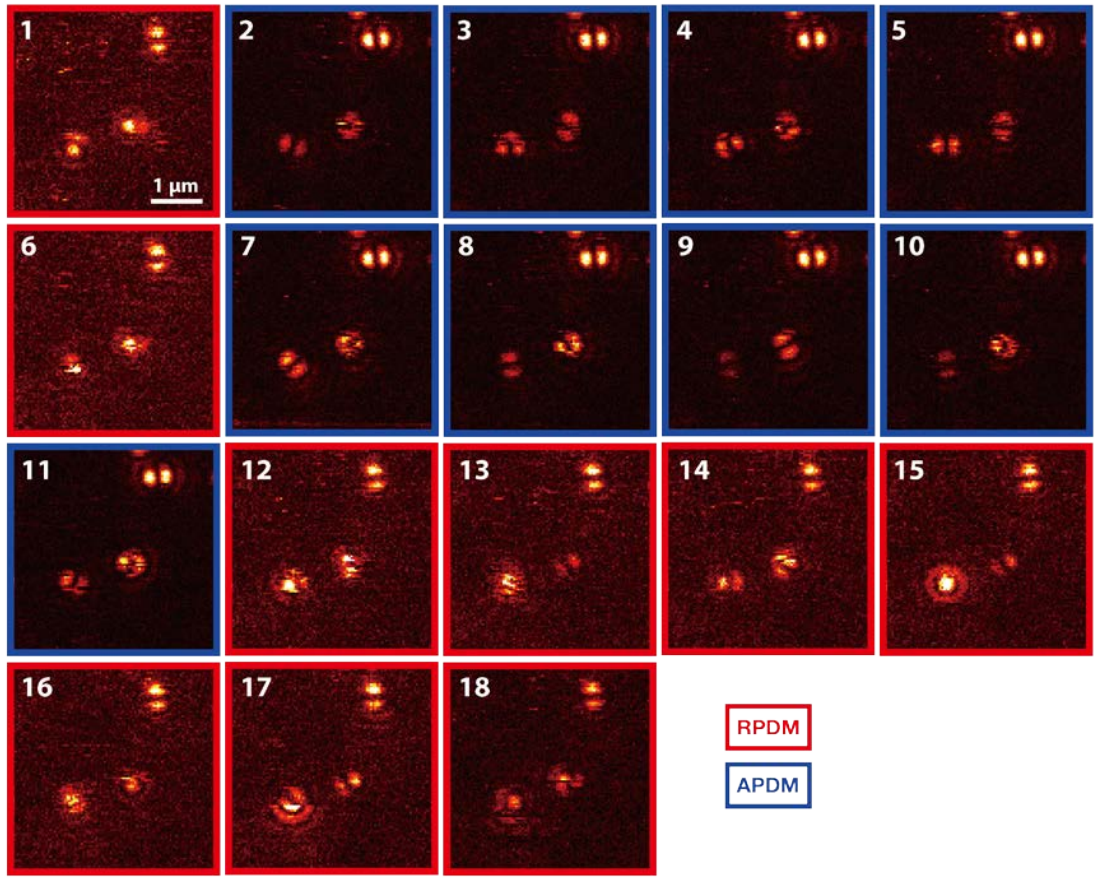

Figure S3. Whole image series of Figure 4 consisting of 18 consecutive scans recorded at same spatial position. The images (1,2,5,7,9,14,15,17) are shown in Figure 4. Images are arranged chronologically. The colored frames show if they are acquired with the APDM (Blue) or RPDM (red). Scanning range is 4 μm × 4 μm (100 pixels × 100 pixels, 5 ms per pixel).

## 5. Successive images recorded from the same single molecule

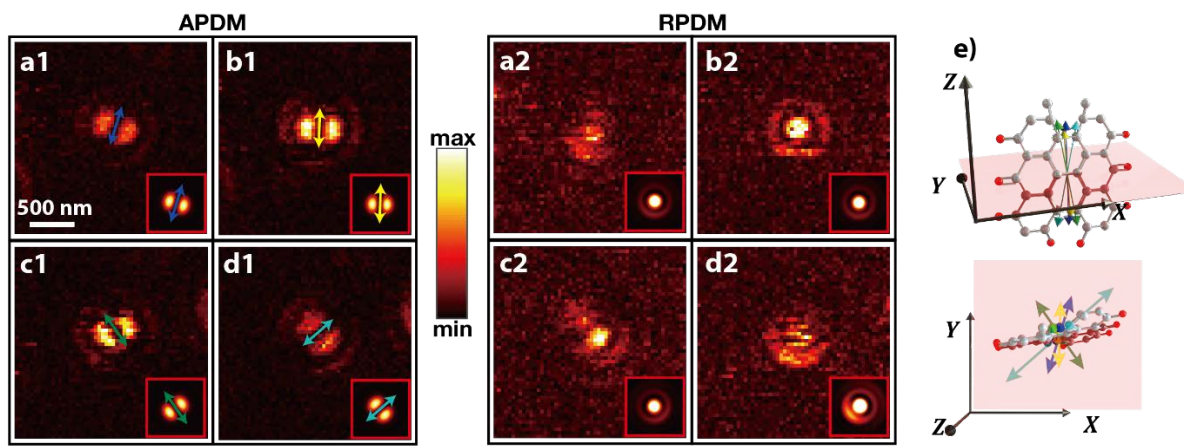

Figure S4. Series of successive images recorded at same spatial position. Eight exemplary images are picked from a total of 25 scans. Scan range is 2 μm × 2 μm (50 pixels × 50 pixels, 5 ms acquisition time per pixel), (a1-d1) are excited with APDM and (a2-d2) are excited with RPDM. Images are arranged in pairs (a1-a2, b1-b2, …) according to the transition dipole of the molecule. The image pattern shows an orientation change in consecutive scans. Inset images in red squares (1.2 μm × 1.2 μm) are simulation results with actual polar angle and azimuthal angle listed in Table S2. Colored arrows (blue, yellow, green, cyan) correspond to the tautomers in Figure 2 and illustrate the orientation of the in-plane dipole moment. (e) Depiction of molecular structure and electronic transition dipoles in the image coordinate system. Top row shows the orientation of the transition dipoles in the image coordinate system. Bottom row illustrates the projection on the X-Y plane.

## 6. Molecules with small polar angle on substrate

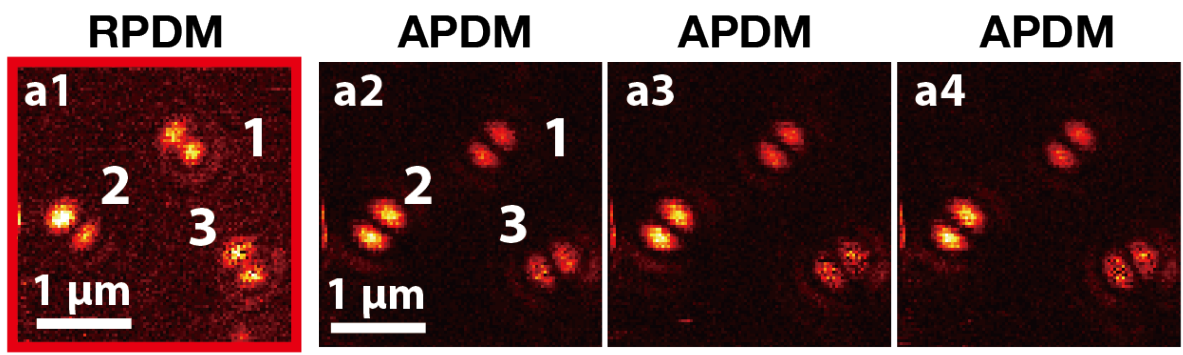

Figure S5 Molecules lie nearly flat on X-Y plane indicated by RPDM excitation patterns in a1. a2-a4 are scanned continuously in 8 minutes and only molecule 3 shows a small angle flipping of the image pattern in a3.

## 7. Calculated azimuthal and polar angle

Table S2 Calculated TDM azimuthal and polar angles of the molecules shown in Figure 4, 5 and S4

|  | Molecule 3 in Fig. 4 | | | | Molecule in Fig. S4 | | | |
|---|---|---|---|---|---|---|---|---|
| Tautomer | $Q_{7,14}$ | $Q_{1,7}$ | $Q_{3,7}$ | $Q_{7,13}$ | $Q_{7,14}$ | $Q_{1,7}$ | $Q_{3,7}$ | $Q_{7,13}$ |
| $\Phi$ | 46.8° | 89.0° | 60.2° | 0.8° | 72.9° | 125.0° | 88.3° | 41.2° |
| $\Theta$ | 167.1° | 166.0° | 167.6° | 160.0° | 11.1° | 11.0° | 10.2° | 22.6° |
|  | Molecule 2 in Fig. 5 | | | | Molecule 3 in Fig. 5 | | | |
| Tautomer | $Q_{7,14}$ | $Q_{1,7}$ | $Q_{3,7}$ | $Q_{7,13}$ | $Q_{7,14}$ | $Q_{1,7}$ | $Q_{3,7}$ | $Q_{7,13}$ |
| $\Phi$ | 131.2° | 174.0° | 152.3° | 50.0° | 94.4° | 58.0° | 79.6° | 166.8° |
| $\Theta$ | 5.8° | 13.1° | 7.5° | 14.0° | 10.2° | 15.8° | 11.4° | 13.6° |